\documentstyle[12pt,axodraw]{article}
\newcommand{\be}{\begin{equation}}
\newcommand{\ee}{\end{equation}}
\newcommand{\bea}{\begin{eqnarray}}
\newcommand{\eea}{\end{eqnarray}}

\newcommand{\gm}{\gamma}
\newcommand{\Gm}{\Gamma}

\newcommand{\ep}{\epsilon}

\newcommand{\dd}{\mbox{d}}

\newcommand{\nn}{\nonumber}
\newcommand{\uk}{\underline{k}}

\begin{document}
\parindent=1.5pc

\begin{titlepage}
\rightline{NPI MSU 98--54/555}
\rightline{hep-ph/9812529}
\rightline{December 1998}
\bigskip
\begin{center}
{{\bf
The strategy of regions for asymptotic expansion of two-loop
vertex Feynman diagrams
} \\
\vglue 5pt
\vglue 1.0cm
{ {\large V.A. Smirnov\footnote{E-mail: smirnov@theory.npi.msu.su.
Supported by the Russian Foundation for Basic
Research, project 98--02--16981 and by Volkswagen Foundation, contract
No.~I/73611.} and
E.R. Rakhmetov\footnote{E-mail: rahmetov@theory.npi.msu.su.
Supported by the Russian Foundation for Basic
Research, project 98--02--16981.}} }\\
\baselineskip=14pt
\vspace{2mm}
{\em Nuclear Physics Institute of Moscow State University}\\
{\em Moscow 119899, Russia}
\vglue 0.8cm
{Abstract}}
\end{center}
\vglue 0.3cm
{\rightskip=3pc
 \leftskip=3pc
\noindent
General prescriptions for evaluation of coefficients at
arbitrary powers and logarithms
in the asymptotic expansion of Feynman diagrams in the Sudakov limit are
discussed and illustrated by two-loop examples. Peculiarities connected
with evaluation of individual terms of the expansion, in particular,
the introduction of auxiliary analytic regularization, are characterized.
\vglue 0.8cm}
\end{titlepage}

{\bf 1.}
The simplest explicit formulae [1--3]
(see brief reviews
in \cite{Sm2}) for the asymptotic expansion of Feynman
diagrams in various off-shell limits of momenta and masses, when
the momenta are considered either large or small in the Euclidean sense,
have been generalized to some typically Minkowskian
on-shell limits [5--7],
in particular, to the Sudakov limit.
The prescriptions for these limits have been formulated using
(pre)subtractions in a certain family of subgraphs
of a given graph.

Recently explicit prescriptions for expanding Feynman integrals
near threshold have been presented \cite{BS}. This was done
with the help of a standard physical strategy based on analysis of
regions in the space of loop momenta.
It should be pointed out, however, that this strategy of regions is
usually applied only for evaluating and summing up the leading logarithms,
in particular, in the Sudakov limit (see. e.g. \cite{Co}).
Note that the information about the leading logarithms is present only
in contributions of some specific regions so that usually one does not
consider integration in other domains.

It was argued (and demonstrated for the threshold expansion)
in \cite{BS} that it is worthwhile
to use this strategy for the evaluation of coefficients at
any power and logarithm for an arbitrary limit.
In such extended form, the strategy reduces to the following prescriptions:

(a) consider all the regions of the loop momenta that are typical for the
given limit and expand, in every region,
the integrand in a Taylor series with respect to the parameters
that are considered small in the given region,

(b) integrate the integrand expanded, in every region in its own way,
over the whole integration domain in the loop momenta,

(c) put to zero any scaleless integral (even if it
is not regulated, e.g., within dimensional regularization).

As it was pointed out in \cite{BS}, it is the step (b) in this procedure
that is far from being trivial. One may believe that this strategy is
legitimate for every limit of momenta and masses. For example, it leads
to the well-known formulae for asymptotic expansions in the case
of typically Euclidean limits \cite{Go,Ch1} (proven in \cite{Sm1})
so that we
have such indirect confirmation at least for them.
Note that, for these limits as well as for the on-shell limit
considered in refs.~\cite{Sm3,ACVS}, the collection of relevant regions is
determined by subdividing all the loop momenta into large (hard) and small
(soft) ones.

In the present paper, we check, by two-loop examples,
this heuristic procedure
for the evaluation of coefficients at arbitrary powers and logarithms
in asymptotic expansions of Feynman diagrams in the Sudakov limit
\cite{Sud}.
We shall consider two commonly accepted variants of this limit
for vertex diagrams with the external momenta $p_1, p_2$ and
$q=p_1-p_2$:

{\em Limit 1.}
Two external momenta are off shell, $p_1^2 =p_2^2 =m^2=-\mu^2$,
$Q^2\equiv -q^2 \to \infty$, all internal masses are
zero.

{\em Limit 2.}
Two external momenta are on shell,  $p_1^2 =p_2^2 =0$;
$Q^2 \to \infty$; some internal masses are non-zero.

We shall calculate the leading power behaviour, including all the logarithms,
$\ln^j (q^2/m^2), \; j=0,1,2,3,4$, of the massless planar diagram
Fig.~1a in the first limit and compare the obtained result
with a known explicit expression \cite{UD}.
After this confirmation
we shall then apply the above heuristic prescription
to the non-planar diagram, Fig.~1b,
(for which no analytical results are known)
in Limit~2 when $m_1=\ldots=m_4=0, \; m_5=m_6=m$.
We shall also use the second example to
describe techniques for evaluation of individual terms of the
expansion. A natural way for evaluation terms with the $1/(m^2)^{2\ep}$
dependence is introduction of an auxiliary analytical regularization.
In contrast to the planar diagram in the second limit where the poles
of the first order in the analytical regularization parameter arise (and
cancel in the sum of two contributions)  \cite{Sm4}, we shall meet, for the
non-planar diagram, poles
up to the second order which are present in five contributions.
These poles are also cancelled in the sum and we obtain a result
which exists within dimensional regularization.

{\bf 2.}
The Feynman integral for Fig.~1a can be written as
\bea
 F_{1}(Q,m,\ep)
= \int  \int \frac{\dd^dk \dd^dl}{(l^2-2 p_1 l+m^2) (l^2-2 p_2 l+m^2)}
\nn \\ \times \frac{1}{(k^2-2 p_1 k+m^2) (k^2-2 p_2 k+m^2) k^2 (k-l)^2} \, .
\label{F1}
\eea
We use dimensional regularization \cite{dimreg} with $d=4-2\ep$.
When presenting our results we shall omit $i\pi^{d/2}$ per loop and,
when writing down separate contributions through  expansion in $\ep$,
we shall also omit $\exp(-\gm_E \ep)$ per loop ($\gm_E$ is the Euler
constant).

Let us choose, for convenience, the external momenta as follows:
\be
p_1 = \tilde{p}_1 +\frac{m^2}{Q^2} \tilde{p}_2 \, , \;
p_2 = \tilde{p}_2 +\frac{m^2}{Q^2} \tilde{p}_1 \, , \;
\tilde{p}_1 = (Q/2,-Q/2,0,0), \;  \tilde{p}_2 = (Q/2,Q/2,0,0)
\ee
so that  $p_i^2 = m^2 \, , \; \tilde{p}_i^2 = 0 \, , \;
2\tilde{p}_1 \tilde{p}_2= 2\tilde{p}_1 p_2 = Q^2$.
In the given limit, the following regions happen to be typical \cite{Co}:
\bea
\label{h}
\mbox{{\em hard} (h):} && k\sim Q\, ,
\nn \\
\label{1c}
\mbox{{\em 1-collinear} (1c):} && k_+\sim Q,\,\,k_-\sim m^2/Q\, ,
\,\, \uk \sim m\,,
\nn \\
\label{2c}
\mbox{{\em 2-collinear} (2c):} && k_-\sim Q,\,\,k_+\sim m^2/Q\, ,
\,\,\uk \sim m \, ,
\nn \\
\label{us}
\mbox{{\em ultrasoft} (us):} && k\sim m^2/Q\, .
\nn
\eea
Here $k_{\pm} =k_0\pm k_1, \; \uk=(k_2,k_3)$. We mean by $k\sim Q$, etc.
that any component of $k_{\mu}$ is of order $Q$.

One should consider any loop momentum $k,l,\ldots$ to be
of one of the above types and allow for various choices of the loop momenta.
(Still it is necessary to avoid double counting.)
Other types of regions give zero contributions, in particular, when
one of the loop momenta is {\em soft}, i.e. $k\sim m$.
However, if some masses of the diagram were non-zero then some
soft regions would generate non-zero contributions (that would start from
a subleading order).

In the leading order, $1/Q^4$, we obtain contributions from the
following nine regions:
(h-h), (1c-h), (2c-h), (1c-1c), (2c-2c), (us-h), (us-1c), (us-2c), (us-us).
In this list, regions for the loop momenta $k$ and $l$ in (\ref{F1})
are indicated in the first and the second place, respectively.

The (h-h) region generates terms obtained by Taylor expanding the integrand
in the expansion parameter, $m$. In the leading order, this is nothing but
the value
of the massless planar diagram at $p_1^2 =p_2^2 =0$ first evaluated
in ref.~\cite{Gons}. Although the result can be expressed in gamma functions
for general $\ep$ with the help of the method of integration by parts
\cite{IBP} (this was first done in
\cite{KL}), we present it here,  for brevity, in expansion in $\ep$
\bea
C^{(1)}_{(h-h)} =
\int  \int \frac{\dd^dk \dd^dl}{(l^2-2 \tilde{p}_1 l) (l^2-2 \tilde{p}_2 l)
(k^2-2 \tilde{p}_1 k) (k^2-2 \tilde{p}_2 k) k^2 (k-l)^2} \,  \\
=   \left(
\frac{1}{4 \ep^4} + \frac{5 \pi^2}{24 \ep^2}  +
  \frac{29 \zeta(3)}{6 \ep} + \frac{3 \pi^4}{32}
\right) \frac{1}{(Q^2)^{2+2\ep}} \, .
\eea

All contributions connected with the ultrasoft regions are easily
evaluated in gamma functions by use of alpha parameters. In the leading
order, we have
\bea
C^{(1)}_{(us-us)}
= \int  \int \frac{\dd^dk \dd^dl}{(-2 \tilde{p}_1 l+m^2)
(-2 \tilde{p}_2 l+m^2)
(-2 \tilde{p}_1 k+m^2) (-2 \tilde{p}_2 k+m^2) k^2 (k-l)^2} \,  \nn\\
= \frac{\Gm(1-\ep)^2 \Gm(2 \ep)^2}{\ep^2 (-m^2)^{4\ep} (Q^2)^{2-2\ep}}
\, ,
\label{us-us}
\\
C^{(1)}_{(us-h)}
= \int  \int \frac{\dd^dk \dd^dl}{(l^2-2 \tilde{p}_1 l) (l^2-2 \tilde{p}_2 l)
(-2 \tilde{p}_1 k+m^2) (-2 \tilde{p}_2 k+m^2) k^2 l^2} \, \nn\\
=
\frac{\Gm(1+\ep) \Gm(1-\ep) \Gm(\ep)^2
\Gm(-\ep)^2}{\Gm(1-2\ep) (-m^2)^{2\ep} (Q^2)^{2}}
\, ,
\label{us-h} \\
C^{(1)}_{(us-1c)}
= \int  \int \frac{\dd^dk \dd^dl}{(-2\tilde{p}_1 l) (l^2-2 p_2 l+m^2)
(-2 \tilde{p}_1 k+m^2) (-2 \tilde{p}_2 k+m^2)}  \nn \\ \times
\frac{1}{k^2(l^2 - (2 \tilde{p}_1 l) (2 \tilde{p}_2 k)/Q^2 )} \, \nn\\
=
\frac{\Gm(1-\ep)^2 \Gm(\ep) \Gm(2\ep)
\Gm(-\ep)}{\ep\Gm(1-2\ep) (-m^2)^{3\ep} (Q^2)^{2-\ep}}
\equiv C^{(1)}_{(us-2c)}  \, .
\label{us-1c}
\eea

Using alpha parameters, the rest contributions can be presented, for
general $\ep$, through Mellin-Barnes integrals
\bea
C^{(1)}_{(1c-1c)}
= \int  \int \frac{\dd^dk \dd^dl}{(-2 \tilde{p}_1 l) (l^2-2 p_2 l+m^2)
(-2 \tilde{p}_1 k) (k^2-2 p_2 k+m^2) k^2 (k-l)^2} \, \nn \\
= \frac{\Gm(\ep) \Gm(-\ep) \Gm(2\ep)}{\Gm(1+\ep) (-m^2)^{2\ep} (Q^2)^{2}}
\nn \\ \times
\frac{1}{2\pi}\int_{-i \infty}^{+i \infty} \dd s \,
\frac{\Gm(s-3\ep) \Gm(s+1-2\ep) \Gm(s+1-\ep)\Gm(\ep-s) \Gm(-s)}{\Gm(s+1-3\ep)}
\nn \\
\equiv C^{(1)}_{(2c-2c)}  \, ,
\label{1c-1c} \\
C^{(1)}_{(1c-h)}
= \int  \int \frac{\dd^dk \dd^dl}{(l^2-2 \tilde{p}_1 l) (l^2-2 \tilde{p}_2 l)
(-2 \tilde{p}_1 k) (k^2-2 p_2 k+m^2)} \nn \\ \times
\frac{1}{k^2(l^2 - (2 \tilde{p}_1 k) (2 \tilde{p}_2 l)/Q^2 )} \,
= \frac{\Gm(\ep) \Gm(-\ep) \Gm(1-\ep)}{\Gm(1-2\ep) (-m^2)^{\ep}
(Q^2)^{2+\ep}}
\nn \\ \times
\frac{1}{2\pi}
\int_{-i \infty}^{+i \infty} \dd s \,
\frac{\Gm(s+1) \Gm(s-\ep) \Gm(s+1+\ep) \Gm(-\ep-s) \Gm(-s)}{\Gm(s+1-2\ep)}
\nn \\
\equiv C^{(1)}_{(2c-h)}  \, .
\label{1c-h}
\eea
We imply the standard way of chosing contours: the UV poles
are to the right and the IR poles to the left of them.
The above Mellin-Barnes integrals are expanded in $\ep$
by shifting the contours and picking up residua at points where
UV and IR poles glue together  when $\ep\to 0$.
As a result we obtain
\bea
(Q^2)^{2} \left[
C^{(1)}_{(1c-1c)}+C^{(1)}_{(2c-2c)}+C^{(1)}_{(1c-h)}+C^{(1)}_{(2c-h)}
\right]
\nn \\ =
 -\frac{1}{2 \ep^4}
+ \left( L^2 - \frac{\pi^2}{2} \right) \frac{1}{2\ep^2}
+ \left(
 \frac{1}{2} L^3- \frac{\pi^2}{6} L   - \frac{17 \zeta(3)}{3}
\right)
\frac{1}{\ep}
+ \frac{7}{24} L^4 - 4 \zeta(3) L - \frac{\pi^4}{144}  \, .
\eea
where $L=\ln (Q^2/\mu^2)$ and we have put $\mu=1$, for brevity.
(Note that, in individual contributions, one has both
$\ln (Q^2/\mu^2)$ and $\ln (\mu^2)$.)

Collecting all nine contributions together we observe
that the poles in $\ep$ which turn out to be of very different
(UV, IR and collinear) nature cancel, with the following result
\be
(Q^2)^2 F_{1} (Q,m,0)
\; \stackrel{\mbox{\small $Q \to \infty$}}{\mbox{\Large$\sim$}} \;
\frac{1}{4} L^4 + \frac{\pi^2}{2} L^2 + \frac{7 \pi^4}{60}  \, ,
\ee
in agreement with the leading order expansion of the well-known explicit
result \cite{UD}.

{\bf 3.}
The expansion of the planar diagram Fig.~1a, with
$m_1=\ldots=m_4=0, \; m_5=m_6=m$, in Limit~2 was obtained in arbitrary
order, following the strategy of subtraction operators,
in \cite{Sm4}.
Note that the same expressions for all contributions of the expansion
can be obtained with the help of the strategy of regions.
The list of non-zero contributions, consists, in this language,
of (h-h), (1c-h), (2c-h), (1c-1c) and (2c-2c) contributions plus
a contribution that starts from the next-to-leading order and comes
from the region
where the momentum of the middle line is soft and the second loop momentum
is considered to be hard.

Let us now consider the expansion of the non-planar diagram, Fig.~1b,
in Limit~2. The Feynman integral can be written as
\bea
 F_{2}(Q,m,\ep)
= \int  \int \frac{\dd^dk \dd^dl}{((k+l)^2-2 p_1 (k+l))
((k+l)^2-2 p_2 (k+l))}
\nn \\ \times
\frac{1}{
(k^2-2 p_1 k) (l^2-2 p_2 l) (k^2-m^2) (l^2-m^2)} \, ,
\label{F2}
\eea
where $p_1$ and $p_2$ satisfy the relations for $\tilde{p}_{1,2}$ in the
previous section.
We shall use as well the second choice of the loop momenta
when $k$ and $l$ are chosen as momenta of lines~3 and~4, respectively,
which is obtained by permutation of the masses and corresponds to (\ref{F2})
with $m_1=m_2=m_5=m_6=0, \; m_3=m_4=m$.

Non-zero contributions to the expansion in the leading order
are generated by the following regions:
(h-h), (h-2c), (2c-h),
(1c-1c), (2c-2c), (2c-1c), (1c-1c)$'$, (2c-2c)$'$ and (us-us)$'$.
As above, we indicate the region for the loop momentum $k$ in the first
place and for $l$ in the second place. We denote the regions for
the second natural choice of the loop momenta by prime.
The (h-h) contribution is given by the massless non-planar diagram.
The result, in expansion in $\ep$, can be found in \cite{Gons}:
\bea
C^{(2)}_{(h-h)}
= \left(
\frac{1}{\ep^4} - \frac{\pi^2}{\ep^2}  -
  \frac{83 \zeta(3)}{3 \ep} - \frac{59 \pi^4}{120}
\right) \frac{1}{(Q^2)^{2+2\ep}} \, .
\label{h-h:NP}
\eea
The (us-us)$'$ contribution is easily evaluated in gamma functions:
\bea
C^{(2)}_{(us-us)'}
= \int  \int \frac{\dd^dk \dd^dl}{(-2 p_1 (k+l))
(-2 p_2 (k+l))
(-2 p_1 k+m^2) (-2 p_2 l+m^2) k^2 l^2} \,  \nn\\
= \frac{1}{(Q^2)^{2-2\ep} (m^2)^{4\ep}}
\left[ \Gm(\ep) \Gm(2\ep) \Gm(1-2\ep)\right]^2 \, .
\label{us-us:NP}
\eea

The (2c-h) contribution is given by
\bea
C^{(2)}_{(2c-h)}
= \int  \int \frac{\dd^dk \dd^dl}{
(l^2-2 p_1 l + (2 \tilde{p}_2 k) (2 \tilde{p}_1 l)/Q^2)
(l^2-2 p_2 (k+l) + (2 \tilde{p}_2 k) (2 \tilde{p}_1 l)/Q^2)}
\nn \\ \times
\frac{1}{
(k^2-2 p_1 k) (l^2-2 p_2 l) (k^2-m^2) l^2} \, ,
\label{2c-h:NP}
\eea
and the same leading order (h-2c) contribution is obtained by
permutation of $k$ and  $l$.
Using alpha parameters and (twice) Mellin-Barnes representation
we obtain
\be
C^{(2)}_{(h-2c)} = C^{(2)}_{(2c-h)} =
\left(
-\frac{3}{\ep^4} + \frac{\pi^2}{\ep^2}  +
  \frac{22 \zeta(3)}{\ep} + \frac{16 \pi^4}{45}
\right) \frac{1}{(Q^2)^{2+\ep} (m^2)^{\ep}} \, .
\label{h-e:NP}
\ee

The (1c-1c) contribution is given by
\bea
C^{(2)}_{(1c-1c)}
= \int  \int \frac{\dd^dk \dd^dl}{(-2 p_1 (k+l))
((k+l)^2-2 p_2 (k+l))}
\nn \\ \times
\frac{1}{
(-2 p_1 k) (l^2-2 p_2 l) (k^2-m^2) (l^2-m^2)} \, ,
\label{1c-1c:NP}
\eea
and the (2c-2c) contribution is obtained by permutation of $k$ and  $l$.
We should also consider similar (1c-1c)$'$ and (2c-2c)$'$
contributions with the second choice
of the loop momenta. The corresponding expressions are obtained
by permutating the masses (see above). The fifth non-zero contribution
of the collinear-collinear type originates from the (2c-1c) region.
It happens that these contributions are regulated dimensionally only
in the sum. It is convenient to introduce an auxiliary analytic
regularization into lines 3 and 4 by
$\frac{1}{(k^2-2 p_1 k)^{1+x_1} (l^2-2 p_2 l)^{1+x_2}}$.
In contrast to the planar two-loop diagram in this limit \cite{Sm3},
we meet, in this example, poles in $x_i$  up to the second order.
In particular, the (2c-1c)
contribution is evaluated in gamma functions, for general $\ep$:
\bea
C^{(1)}_{(2c-1c)}
= \int  \int \frac{\dd^dk \dd^dl}{
(-2 p_1 l + (2 p_2 k) (2 p_1 l)/Q^2)
(-2 p_2 k + (2 p_2 k) (2 p_1 l)/Q^2)}
\nn \\ \times
\frac{1}{(k^2-2 p_1 k) (l^2-2 p_2 l) (k^2-m^2) (l^2-m^2)} \, ,
\nn \\
= \frac{\Gm(x_1)\Gm(x_2)\Gm(-x_1-\ep)\Gm(-x_2-\ep)\Gm(x_1+\ep)\Gm(x_2+\ep)
}{\Gm(1+x_1) \Gm(1+x_2)\Gm(-\ep)^2 (-m^2)^{x_1+x_2+2\ep} (Q^2)^{2}}
\, ,
\label{2c-1c:NP}
\eea
Using the technique of alpha parameters and Mellin-Barnes representation
for other four (e-e) contributions, we obtain, for each of them,
a result in expansion in $x_i$.
Then we switch off the analytic regularization
(first, $x_2\to x_1$ and then $x_1\to 0$),
observe that, in the sum of all the five contributions,
the singular dependence
in $x_i$ drops out and obtain the following result in expansion in $\ep$:
\bea
(Q^2)^{2} \left[
C^{(2)}_{(1c-1c)}+C^{(2)}_{(2c-2c)}+C^{(2)}_{(1c-1c)'}+C^{(2)}_{(2c-2c)'}
+ C^{(2)}_{(2c-1c)} \right]
\nn \\
= \frac{19}{4 \ep^4}
- \frac{9}{2\ep^3} L+
\left(L^2- \frac{11\pi^2}{4} \right)  \frac{1}{2 \ep^2}
- \left( \frac{3\pi^2}{4} L + \frac{97\zeta(3)}{6}\right) \frac{1}{\ep}
\nn \\
 + \frac{\pi^2}{12} L^2 + 9 \zeta(3) L - \frac{23\pi^4}{32} \, ,
\label{e-eRes:NP}
\eea
where $L=\ln (Q^2/m^2)$ and we have put $m=1$, for brevity.

Collecting all the leading order contributions we see that the poles
in $\ep$ are canceled and we arrive at the following result:
\be
(Q^2)^2 F_{2} (Q,m,0)
\; \stackrel{\mbox{\small $Q \to \infty$}}{\mbox{\Large$\sim$}} \;
\frac{7}{12} L^4 - \frac{\pi^2}{2} L^2 + 20 \zeta(3) L
- \frac{31 \pi^4}{180}  \, .
\ee

On the expense of a computer algebra, it is possible to
extend this result to any order in $1/Q^2$.

\vspace{2mm}

{\em Acknowledgments.}
V.S. is grateful to M.~Beneke, K.G.~Chetyrkin and A.I.~Davydychev
for useful discussions.

\vspace{1cm}

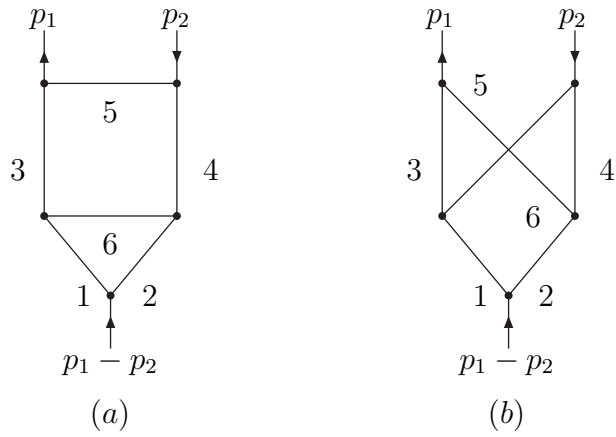
\begin{figure}[thb]
\begin{picture}(400,160)(-100,0)

\Line(10,100)(60,100)
\Line(10,150)(60,150)
\Line(10,100)(10,150)
\Line(60,150)(60,100)
\Line(60,100)(35,70)
\Line(35,70)(10,100)
\ArrowLine(35,50)(35,70)
\ArrowLine(10,150)(10,170)
\ArrowLine(60,170)(60,150)


\Text(35,45)[]{$p_1-p_2$}

\Text(10,175)[]{$p_1$}
\Text(60,175)[]{$p_2$}
\Text(35,25)[]{$(a)$}
\Text(  0,118)[]{3}
\Text( 25, 70)[]{1}
\Text( 73,118)[]{4}
\Text( 50, 70)[]{2}
\Text( 35,140)[]{5}
\Text( 35, 90)[]{6}

\Vertex(10,100){1.5}
\Vertex(60,100){1.5}
\Vertex(10,150){1.5}
\Vertex(60,150){1.5}
\Vertex(35,70){1.5}


\Line(160,100)(210,150)
\Line(160,150)(210,100)
\Line(160,100)(160,150)
\Line(210,150)(210,100)
\Line(210,100)(185,70)
\Line(185,70)(160,100)

\Vertex(160,100){1.5}
\Vertex(210,100){1.5}
\Vertex(160,150){1.5}
\Vertex(210,150){1.5}
\Vertex(185,70){1.5}

\ArrowLine(185,50)(185,70)
\ArrowLine(160,150)(160,170)
\ArrowLine(210,170)(210,150)

\Text(185,45)[]{$p_1-p_2$}
\Text(160,175)[]{$p_1$}
\Text(210,175)[]{$p_2$}
\Text(150,118)[]{3}
\Text(175, 70)[]{1}
\Text( 223,118)[]{4}
\Text( 200, 70)[]{2}
\Text( 175,150)[]{5}
\Text( 195, 100)[]{6}

\Text(185,25)[]{$(b)$}
\end{picture}
\vspace{-10pt}
\caption {
(a) Two-loop planar vertex diagram.
(b) Two-loop non-planar vertex diagram. }
\label{2l}
\end{figure}
\end{document}